# Review on vortex dynamics in the left ventricle as an early diagnosis marker for heart diseases and its treatment outcomes

Mahesh S. Nagargoje[*1, 3], Eneko Lazpita[1],

Jesús Garicano-Mena[1,2] and Soledad Le Clainche[1,2]

[1] ETSI Aeronáutica y del Espacio - Universidad Politécnica de Madrid, 28040 Madrid, Spain

[2] Center for Computational Simulation (CCS), 28660 Boadilla del Monte, Spain

[3] LaBS-CompBiomech, Politecnico di Milano, Milan, Italy

*Corresponding author. E-mail: msnagargoje@gmail.com

## Abstract

The heart is the central part of the cardiovascular network. Its role is to pump blood to various body organs. Many cardiovascular diseases occur due to an abnormal functioning of the heart. A diseased heart leads to severe complications and in some cases death of an individual. The medical community believes that early diagnosis and treatment of heart diseases can be controlled by referring to numerical simulations of image-based heart models. Computational Fluid Dynamics (CFD) is a commonly used tool for patient-specific simulations in the cardiac flows, and it can be equipped to allow a better understanding of flow patterns. In this paper, we review the progress of CFD tools to understand the flow patterns in healthy and dilated cardiomyopathic (DCM) left ventricles (LV). The formation of an asymmetric vortex in a healthy LV shows an efficient way of blood transport. The vortex pattern changes before any change in the geometry of LV is noticeable. This flow change can be used as a marker of DCM progression. We can conclude that understanding vortex dynamics in LV using various vortex indexes coupled with data-driven approaches can be used as an early diagnosis tool and improvement in DCM treatment.

## Graphical Abstract

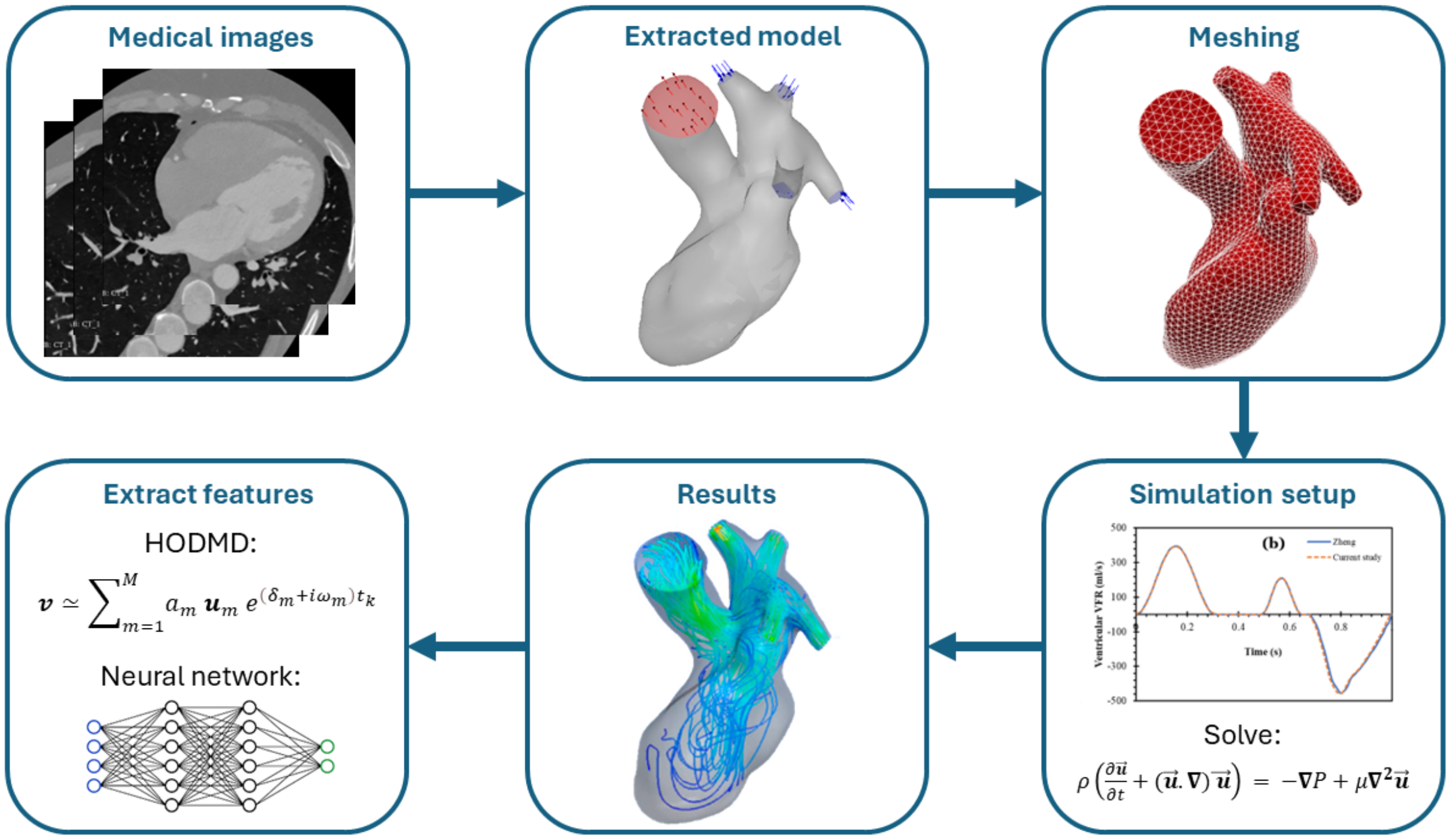

## Impact Statement

This manuscript discusses recent advances in vortex dynamics in the left ventricle and its role in heart disease progression and early diagnosis. Various vortex formation indexes have been formulated in the past and these can be useful for the early diagnosis of flow pattern changes and better treatment management. The change in flow patterns due to structural change in the left ventricle has been critically analyzed by discussing hemodynamics in healthy and dilated cardiomyopathy models. Challenges, opportunities, and future directions for left ventricle hemodynamics have been discussed.

## 1.0 Introduction

Cardiovascular diseases (CVDs) are a leading cause of death globally. According to the World Health Organization (WHO), CVDs account for approximately 31% of all deaths worldwide. In the United States, heart disease and stroke are major contributors to this burden, accounting for more deaths than all forms of cancer and Chronic Lower Respiratory Disease (CLRD) combined (Kaptoge *et al.* 2019; Virani *et al.* 2021). These diseases include a range of conditions such as coronary heart disease, stroke, and heart failure, and are caused by a variety of factors including high blood pressure, high cholesterol, smoking, physical inactivity, and unhealthy diet. Heart attacks is the major cause of death in CVDs, which account for 25% of deaths to total deaths. Disabilities due to CVDs can have a significant impact on individuals, and communities, and definitely lead to decreased quality of life and increased healthcare costs. Understanding the global burden of heart diseases and their risk factors is crucial for developing effective prevention and treatment strategies.

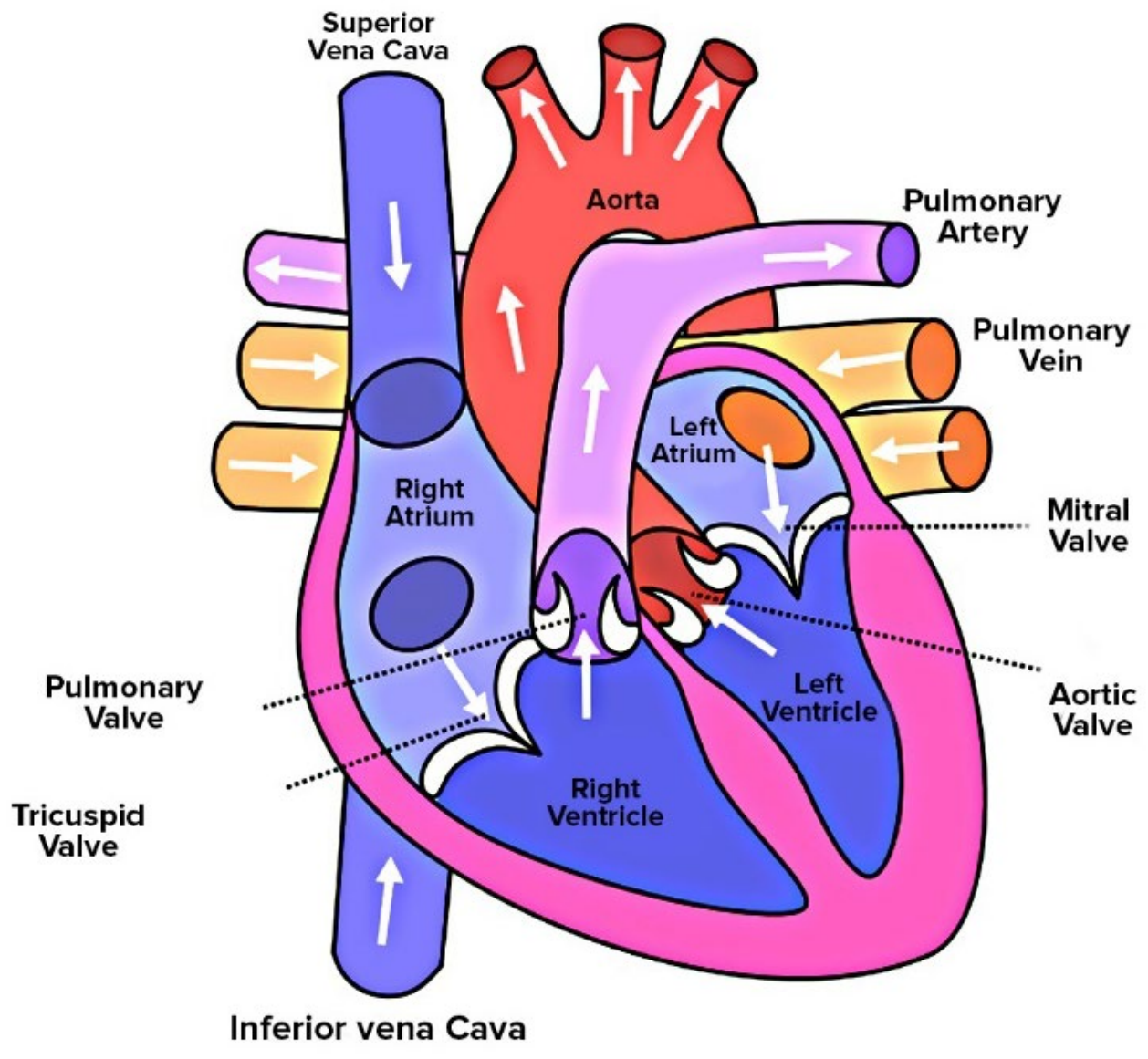

Figure. 1. Schematic of human heart consisting of four chambers such as right atrium, left atrium, right ventricle, and left ventricle. Heart valves controlling blood flow between chambers are shown in whole heart model. Schematic is reused under GNU free documentation license from Free Software Foundation (Human Heart 2023).

The human heart consists of four chambers, a pair of atriums and ventricles, as shown in Fig. 1. Atria are used for collecting blood whereas ventricles for pumping the blood. The right atrium collects the deoxygenated blood from various organs and the left atrium collects oxygenated blood from lungs via pulmonary veins. The right ventricle pumps the blood to the lungs for oxygenation and the left ventricle pumps the oxygenated blood to the rest of the body. The left ventricle (LV) plays a crucial role in the functioning of the heart, being responsible for supplying blood to all the tissues throughout the body. LV collects the oxygenated blood from left atrium during diastolic filling and ejects the blood during systole. The vortex patterns during diastolic filling differentiates between normal and abnormal left ventricle flow dynamics. Asymmetric and smooth vortex formation in the LV is considered as a healthy way of blood transport. However, the presence of an unnatural vortex in the left ventricle (LV) is often associated with cardiac diseases. One of the very common LV diseases is dilated cardiomyopathy (DCM), which is a condition in which the LV chamber becomes enlarged, with thinner and weaker walls, which corresponds to decreased cardiac function (Grossman *et al.* 1974; Jefferies and Towbin 2010). Due to change in anatomy of the LV, the flow and vortex patterns changes. It has been observed that DCM patients have less ordered blood flow patterns with decreased vortex strength, decrease in flow propagation velocity, and more stagnation at apex of LV (Baccani *et al.* 2002b; Loerakker *et al.* 2008). Changes in blood flow pattern have been analyzed using computational fluid dynamics (CFD), by investigating various flow and vortex indexes.

Vortex patterns in the left ventricle are crucial for efficient blood transport and have been studied extensively in relation to cardiovascular disease. Past studies have shown that abnormal flow patterns due to vortex breakdowns can lead to energy dissipation and decreased blood ejection efficiency. Decrease in ejection fraction increases the in-turn risk for heart failure (Dabiri 2009; Pedrizzetti and Domenichini 2005, 2015; Pedrizzetti and Sengupta 2015). The vortex in a healthy LV is asymmetric and non-planar which avoid eddies formation by recirculating and propagating towards outlet without colliding with each other (Kilner *et al.* 2000). The asymmetric donut-shaped vortex ring formation occurs during E-wave diastolic filling. An index has been formulated to characterize the vortex ring formation quantitatively and its propagation: the *vortex formation number* (*VFN*). VFN is a measure of the length to diameter ratio of the ejected fluid, which is directly proportional to ejection velocity and inversely proportional to orifice opening. A vortex formation number between 3-5 has been considered as optimal, whereas values beyond 5 leads to instability and excessive energy dissipation (Gharib *et al.* 2006). Recently, the vortex in LV has been used as an early predictor of cardiovascular outcomes. It is believed that abnormal flow patterns in LV could signal the presence of heart malfunction even before noticeable structural changes are evident (Pedrizzetti *et al.* 2014).

Numerical studies have become an essential tool in understanding the hemodynamics and vortex patterns of the left ventricle since last couple of decades. CFD has been used to study the blood flow patterns in the left ventricle, providing insights into the vortex dynamics in both healthy and diseased hearts (Chan *et al.* 2013b; Doost *et al.* 2016a). The CFD studies were started with 2D models of LV and analyzed the formation of vortex in healthy and diseased LV and found that the flow wave propagation velocity reduced in dilated LV (Baccani *et al.* 2002a, 2002b; Chan *et al.* 2013c; Loerakker *et al.* 2008; Pierrakos and Vlachos 2006; Vierendeels *et al.* 2000). Further, numerical simulation of three-dimensional ideal healthy LV models found asymmetric vortex generation (Schenkel *et al.* 2009). Recently, high-resolution imaging techniques such as MRI have been combined with CFD to obtain detailed flow characteristics and improve our understanding of the complex hemodynamics in the left ventricle (Khalafvand *et al.* 2015; Labbio and Kadem 2018). Past studies investigated that the effect of inlet waveform specified at pulmonary veins does not affect flow patterns in patient-specific LV and stagnation zones at the apex of LV are prone to thrombi formation (Lantz *et al.* 2019; Liao *et al.* 2016). Most of the studies have the major limitation of considering walls of LV as rigid, which is not close to in vivo situation. Realistic studies have modeled the contraction and expansion of LV using a moving mesh approach. Consideration of moving walls improves the accuracy of the vortex pattern prediction and the hemodynamics in LV (Chan *et al.* 2013a, 2019; Cheng *et al.* 2005; Vedula *et al.* 2014). Recently published articles considered the coupling of mitral valve (MV) movement with LV hemodynamics and found that it helps in risk stratification and optimization of heart therapies (Gao *et al.* 2017).

Despite the complexity of cardiac flows, this review will focus on showing an extensive review on the formation of the vortex. It should be acknowledged that factors such as mitral and aortic valve modeling, patient conditions, and disease alterations influence the flow. However, we will specifically examine the literature on the vortical patterns in these flows, setting aside other phenomena beyond the scope of this paper.

This review, which includes over 129 references, summarizes significant advances from the last two decades in the fundamentals of cardiac flows from a numerical perspective. It also compares these with experimental or medical data. The selected publications provide a comprehensive overview of the current technology for studying cardiac flows, offering new insights into the flow physics of healthy and diseased hearts. Additionally, this review highlights the need for validating optimal and abnormal vortex patterns using extensive datasets of experimental data and clinical observations. The review also considers the potential of data-driven methods, including machine learning, for identifying abnormal vortex patterns as markers for heart function monitoring. Trained machine learning models could become invaluable for assessing early outcomes of heart valve surgeries and indicating the success of various heart procedures. This opens new discussions on the condition selection necessary for accurate CFD simulations of the left ventricle.

## 2.0 Vortex dynamics in the left ventricle

The filling of blood in the left ventricle occurs due to the mitral valve opening and flowing of blood into the left ventricle through the small MV orifice. The mitral valve consists of two triangular cusps of unequal size, they are stronger in mechanical strength than tricuspid valves. The expansion of the left ventricle during early diastolic flow (during the E-wave) creates a reduced pressure in the LV, which generates a gradient for blood flow from mitral leaflets (high pressure) to left ventricle apex (low pressure). Decreased pressure thus generated in the LV chamber induces the mitral cusps to open. The filling of blood from the left atrium is observed while low pressure exists in the LV. Atrium contraction also contributes to LV filling which is called A-wave filling. The blood flow through a healthy mitral valve is generally laminar: this promotes a smooth transfer of blood from the MV to aortic valve (AV) through LV, as shown in Fig. 2(a). The smooth transfer of blood is observed from generation of a stable asymmetric vortex.

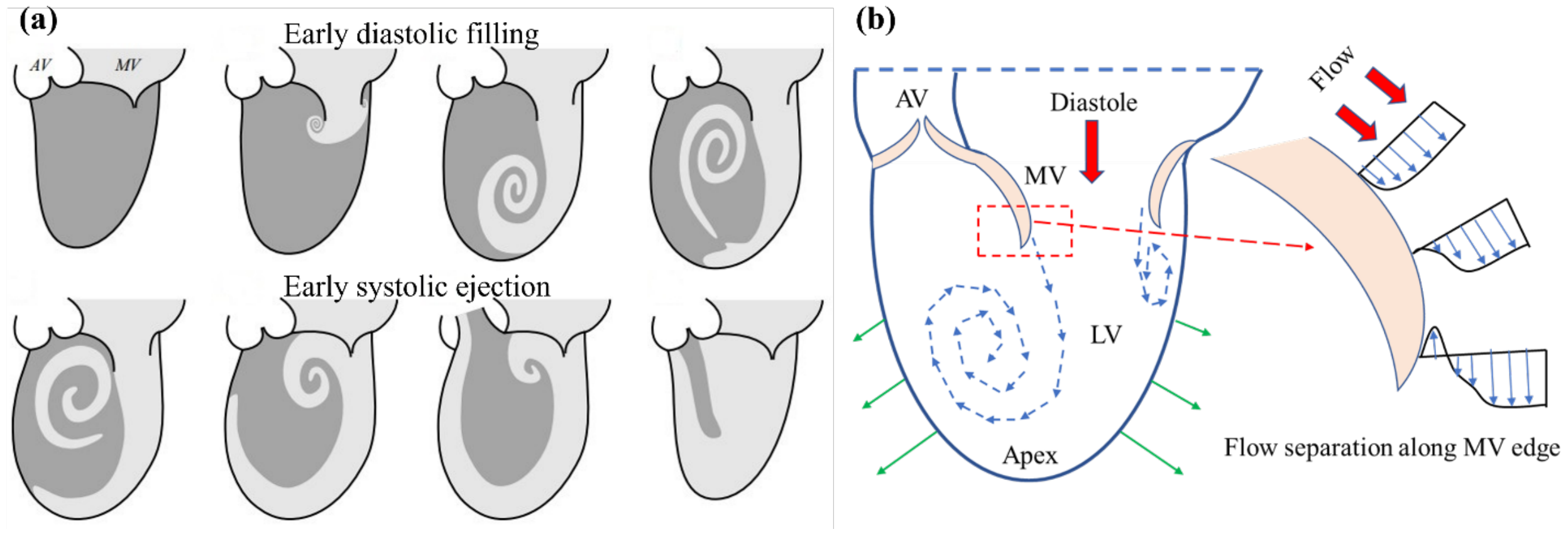

Fig. 2. (a) Schematic of blood transport during filling and ejection of blood through a healthy LV, dark grey: old blood, and light grey: new blood. Adapted with permission from a previous numerical study (Di Labbio *et al.* 2022) (b) Blood flows (thick red arrow) into the left ventricle through the MV, blood develops two shear layers of different velocities at the trailing edge of MV. The boundary layer separation leads to an adverse pressure gradient and blood stream rolls up into a vortex.

The boundary layer separation of MV leaflet's trailing edge and the subsequent development of shear layers is the triggering point of vortex formation. During blood flow through the mitral valve, blood develops shear layers of increasing velocity away from valve edge, as shown in Fig. 2(b). The velocity of the blood layer close to valve edge is almost zero and away from the valve edge shows a higher velocity. Due to the curvature of MV edges, flow separation occurs and due to gradient in velocity shear layers, the bending of the streamlines induces a rolling motion of blood, and the subsequent formation of a vortex on both sides of MV. During roll-up process, the distance between adjacent vortex turns reduces and the vortex cone forms a conical like flow trajectory. The base of a cone can be referred to as *initial vortex size*, whereas the tapered end of cone is considered as the *end* of the vortex helical path. The vortex formation in LV is smoother (in contrast to chaotic or turbulent flow) phenomena due to confined and limited expansion of left ventricle, which restricts the vortex size to grow. The formation of vortex stores the incoming translational kinetic energy into a rotary motion. Ill-functioning LVs show abnormal vortex formation flows and it may lead to turbulence and excessive energy dissipation. It has been reported that malfunctioning of the heart typically results in changes in vortex dynamics or size and number of vortices.

The vortex formed in a healthy LV is asymmetric: these asymmetries been proven beneficial for smooth transfer of blood from left ventricle to aorta. This asymmetric shape is due to the unbalanced shape of mitral valve, since the anterior leaflet (Fig. 2b left cusp of MV) is larger than posterior leaflet (right cusp of MV). Also, the area available for flow and fluid wall interaction is different for both the leaflets. The posterior leaflet has a free edge and more area to flow while anterior leaflet finds the LV wall adjacent to the MV leaflet. The generated vortex at MV propagates into LV until it gets pinched-off from the flowing jet at the end of diastolic phase. The vortex can be visualized as the donut shaped ring in Fig. 3. The asymmetry in the vortex size is also reflected in vortex ring thickness. The ring thickness is lower on the posterior side as compared with anterior. The ring thickness depends on leaflet lengths and non-homogenous pressure gradient in the LV. The vortex ring rotates while propagating towards the LV apex. Moreover, the ring-shaped vortex changes its axis of rotation by ninety degrees and efficiently aligns itself in the flow direction towards the aorta. A lower thickness ring shifts towards LV apex by rotation and can be an efficient way of momentum transfer, as shown in Fig. 3 (Kheradvar *et al.* 2012).

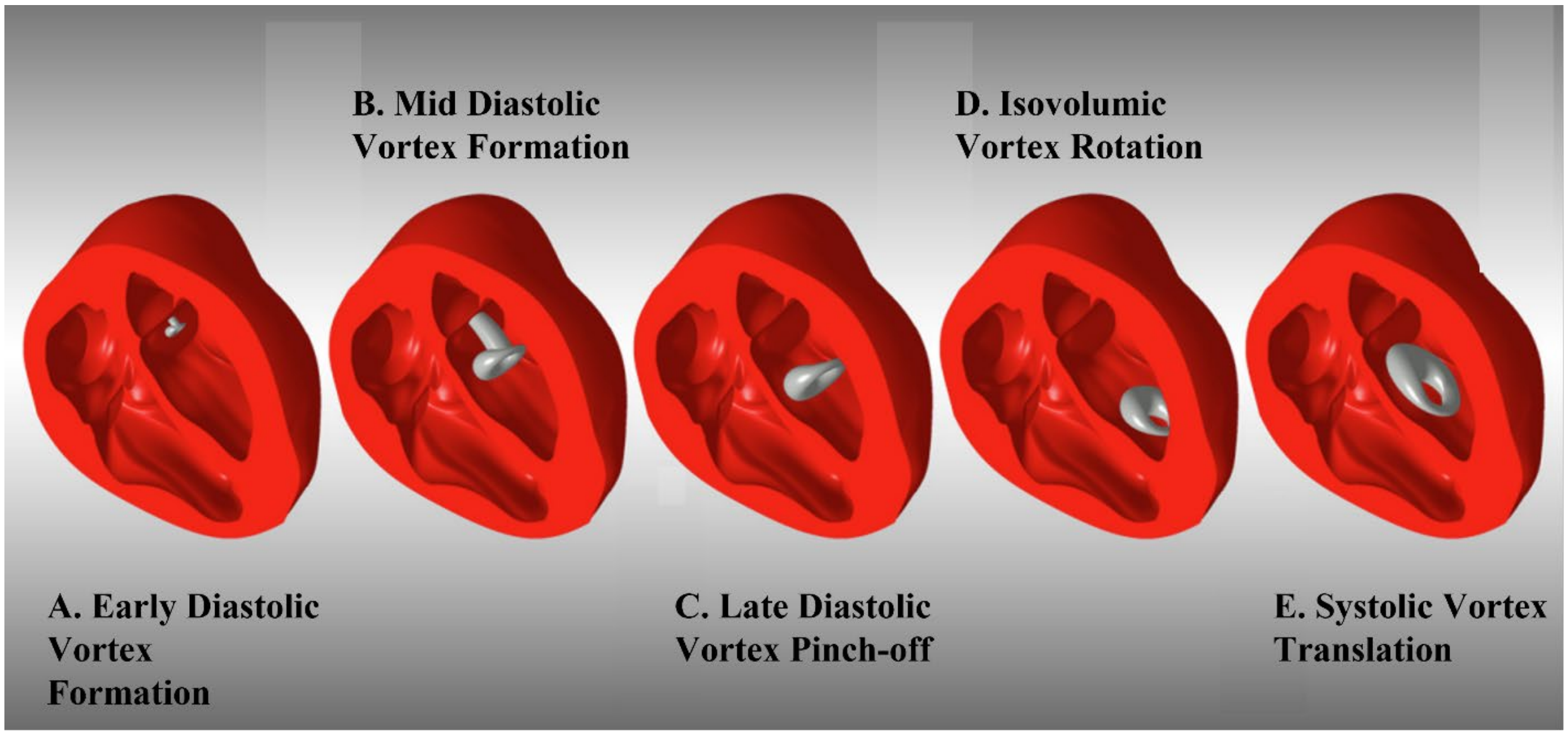

Fig. 3. Vortex formation in the healthy LV and the vortex ring pinch-off and rotation during a cardiac cycle. Adapted with permission from a previously published article (Kheradvar *et al.* 2012).

Figure 4 shows the intraventricular pressure distribution during LV filling in a healthy subject. Before E-wave peak, the pressure gradient points from MV orifice towards apex and the flow is driven in the same direction. The generation of a lower pressure zone at the apex is the result due to LV expansion. After the peak velocity has been attained, the pressure distribution reverses until the apical pressure exceeds in comparison with base. This pressure gradient directs the flow towards the middle cavity of LV. During diastasis the pressure gradient does not significantly vary in the LV cavity. Again, the pressure gradient from base to apex is generated due to the contraction of atrial chamber before A-wave peak. Finally, the contraction of the LV results in pressure gradient from apex to LV outlet and blood smoothly transfers towards aorta in the form of stable orderly manner converging vortices.

Recall that, LV disfunction and abnormal vortex patterns are strongly connected. The difference between efficient and inefficient LV ejection can be understood and even quantified more easily by introducing several vortex formation indexes. The vortex formation number (VFN) is used to define the optimal range of blood transport and its optimal range is in between 3.5-5.5. It can be expressed as given in equation 1:

$$VFN = \frac{\bar{U}}{\bar{D}} . T \qquad (1)$$

where $\bar{U}$ is the time-average velocity of blood stream coming out from MV, $\bar{D}$ is the time average mitral orifice diameter, and $T$ is the duration of diastolic filling during E-wave.

The VFN range between 3.5-5.5 has been observed as an efficient way of fluid transport. Indeed, all biological propulsions observed in nature are shown a VFN between 3.5-5.5 (Dabiri 2009). If the range goes beyond 5.5 the vortices are formed as a trailing jet, which is inefficient (Gharib *et al.* 1998). The fluid transport by vortices is more efficient than by a steady jet or trailing jet (Krueger and Gharib 2003; Xiang *et al.* 2018). The vortex ring will continue to grow until VFN reaches 4. For VFNs greater than 4 result in vortex instability which in turn prevents vortex growth on the grounds of energetic constraints: this leads to the trailing jet regime (Dabiri and Gharib 2004; Gharib *et al.* 1998). A single vortex ring is formed when the VFN is less than or within the optimal range. For example, mitral stenosis reduces the area available to flow and increases the blood stream velocity, which leads to an increase in VFN beyond optimal range. In contrast, dilated

LV decreases VFN below 3.5 which results in reduction in ejection fraction (EF). The EF can be expressed in terms of VFN as in equation 2.

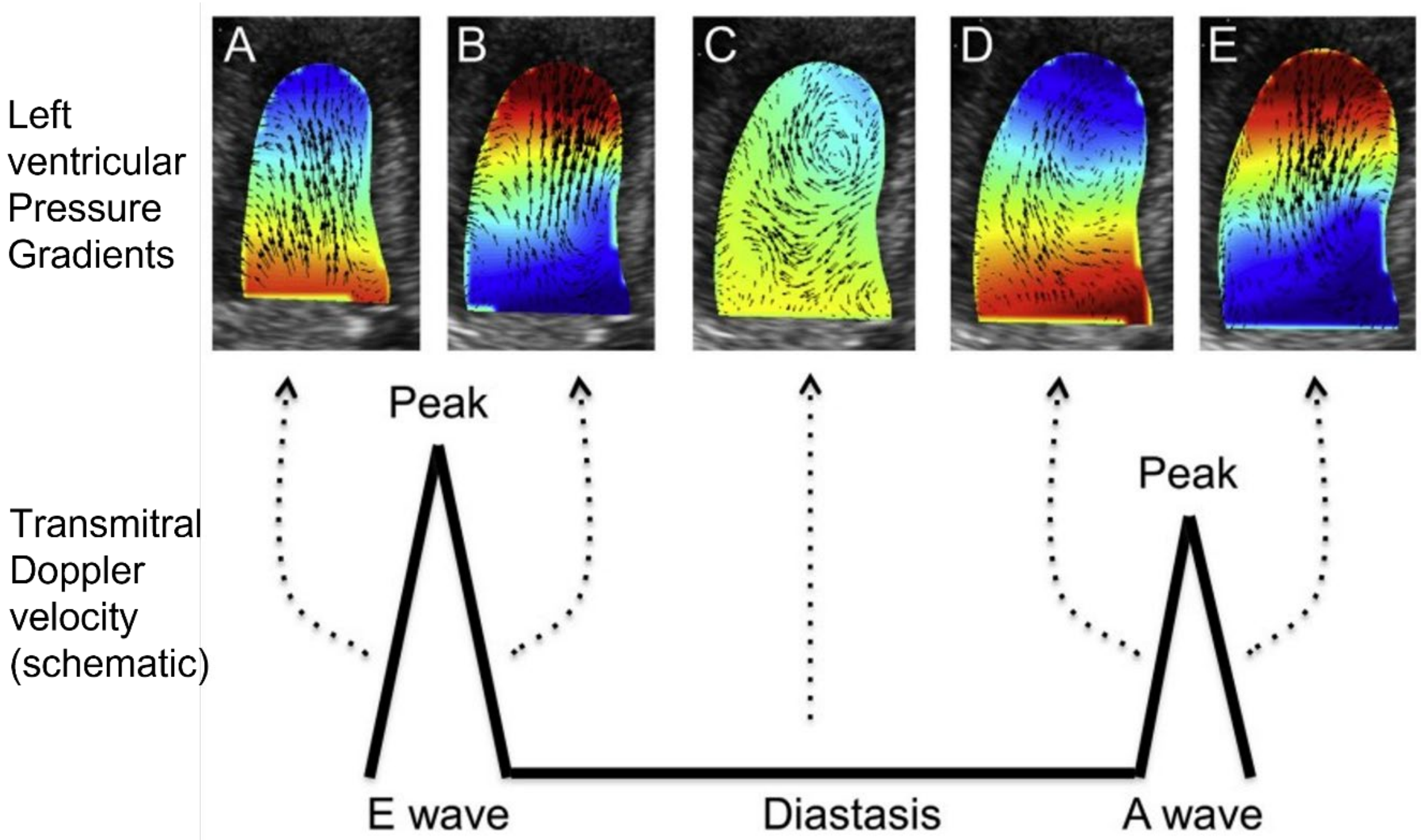

Fig. 4. Intraventricular pressure distribution in LV during diastolic filling phase in a healthy subject. (a) Pressure gradient increases from base to apex and flow accelerates accordingly during early diastole before E-wave peak; (b) Pressure gradient is directed from apex to base and inflow decelerates during end of E-wave; (c) No significant pressure gradient occurs during diastasis; (d) Pressure gradient increases from base to apex and flow accelerates accordingly during beginning of atrial systole before A-wave peak; (e) Pressure gradient is directed from apex to base and accelerating flow towards LV outlet by smooth vortex rotation. Blue: lower, and red: higher values. Adapted with permission from a previously published article (Mele *et al.* 2018).

$$VFN = \frac{4(1-\beta)}{\pi} \cdot \alpha^3 \cdot EF, \qquad (2)$$

where $\beta$ is the fraction of stroke volume in the LV obtained from left atrium (LA) and can be expressed as equation 3.

$$\beta = \frac{V_A}{EDV} = \frac{VTI_A \times \frac{\pi}{4} D_E^2}{EDV}, \tag{3}$$

where $V_A$ is the blood volume entered LV during atrial contraction, $EDV$ is the LV end diastolic volume, $VTI_A$ is the velocity-time integral of A-wave, $D_E$ is the effective diameter of mitral orifice area.

The variable $\alpha$ is a purely geometric parameter of the LV, defined by:

$$\alpha = \frac{EDV^{\frac{1}{3}}}{\bar{D}} \tag{4}$$

where $EDV$ is the LV end-diastolic volume and $\bar{D}$ is the time-averaged mitral valve diameter defined in equation 1.

Comparison of intraventricular flow fields in healthy and DCM patients shows significant differences in velocity profiles, three-dimensional flow fields, and pressure distribution, as shown in Fig. 5 (Mangual *et al.* 2013). The flow patterns are shown at the end of E-wave diastolic phase, and provide an important understanding of flow patterns in healthy and DCM patients. In the healthy left ventricle, anticlockwise rotation of vortex dominates the central cavity of LV. However, diseased (DCM) LVs show a more complex vortex pattern at the middle of LV (Fig. 5: bottom). The three-dimensional flow structures visualized by $\lambda_2$ criterion differentiates that the vortex is more stretched towards apex in the healthy case as compared to more compact vortex in a DCM patient. The vortical circulation is restricted to the central cavity in DCM patient, while it is well distributed in healthy subjects. The vortex strength is higher in healthy cases as compared with DCM patients, as shown in Fig. 5. The vortex core is very compact and limited to central LV cavity for DCM patients. The pressure difference between base and apex is also significantly different in healthy and dilated case, as shown in Fig. 5(right). The pressure gradient in dilated LV is lower, which leads to lower ejection fraction. Note also that the presence of stagnation regions is prone to thrombi formation and leads to dangerous situations. A thrombus may propagate through the arteries and cut off blood supply in narrower arteries leading to further complications. In summary, a well distributed blood flow throughout LV is indicative of healthy flow pattern.

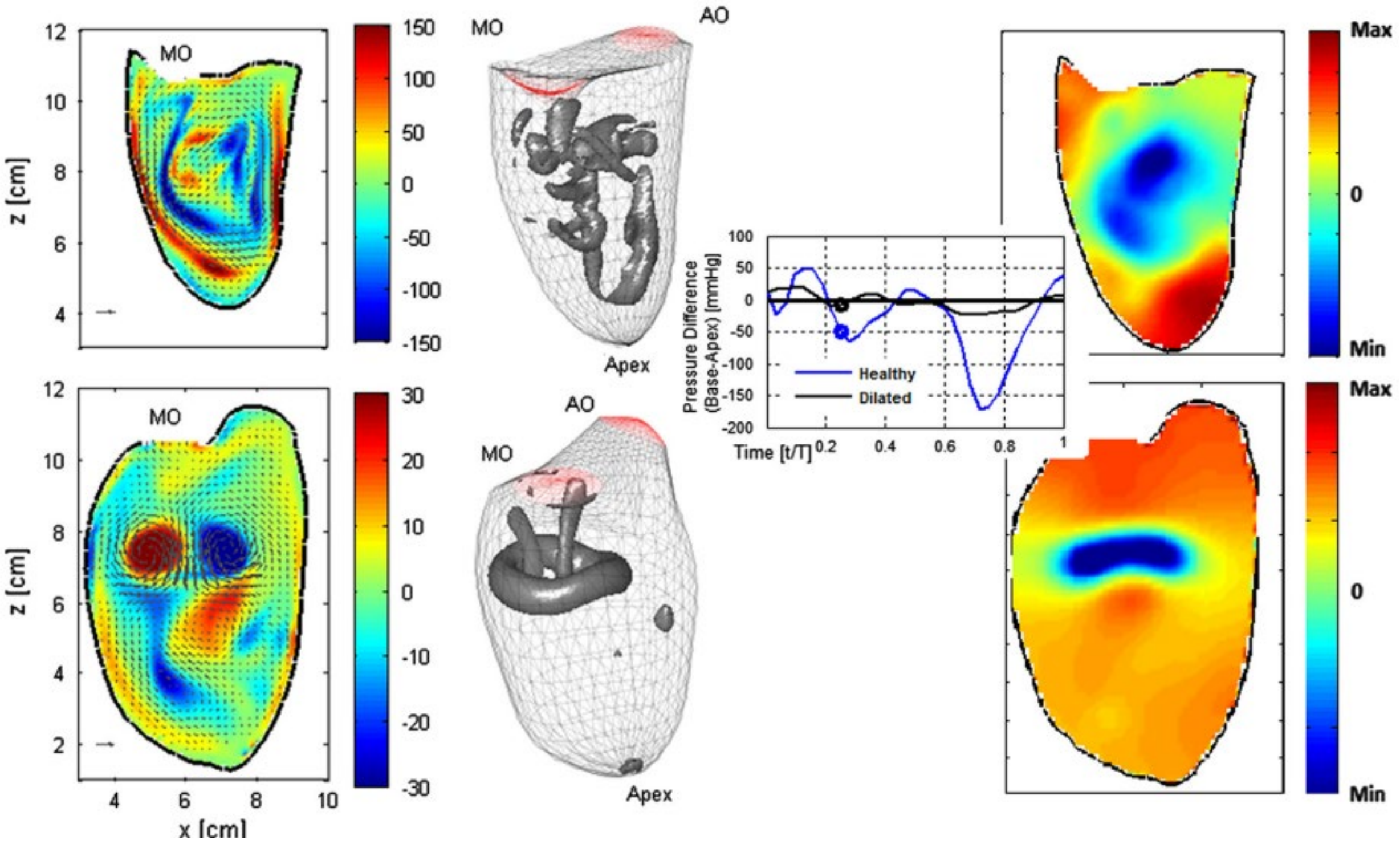

Fig. 5. Numerical simulations for diastolic intraventricular flow at the end of E-wave for a healthy (top) and DCM patient (bottom). Comparison of flow structures for healthy and DCM patient, left: velocity vectors at midplane of LV; middle: three-dimensional vortex fields by iso-surface of $\lambda_2$ criteria; and right: pressure distribution at the midplane. Adapted with permission from previously published numerical simulations (Mangual *et al.* 2013).

Various studies simulated healthy and diseased LV models and compared vortex patterns to understand the changes in vortex dynamics. Patient-specific LV models have been simulated by considering resistance and compliance/flexibility of arteries. Healthy and DCM LV models have been simulated and compared the vortex patterns before and after surgery of DCM LV models, as shown in Fig. 6 (Doenst *et al.* 2009). In a healthy LV, the asymmetric vortex is formed just after the mitral valve opening which was shown by CFD (Fig. 6). The asymmetry might help to minimize the dissipative interaction between inflow, vortical structures, and outflow (Kilner *et al.* 2000). The vortex ring is pinched-off and rotated in clockwise direction to redirect the flow towards aorta. In the dilated LV (pre-operative case), dilation at the apex of LV deforms the vortex towards dilated regions which causes momentum loss during diastole. Similar flow structures were observed during systole for healthy and DCM cases. After the removal of dilated region at LV apex, the vortex deformation was reduced and blood flowed more efficiently towards aorta with lower dissipation or losses.

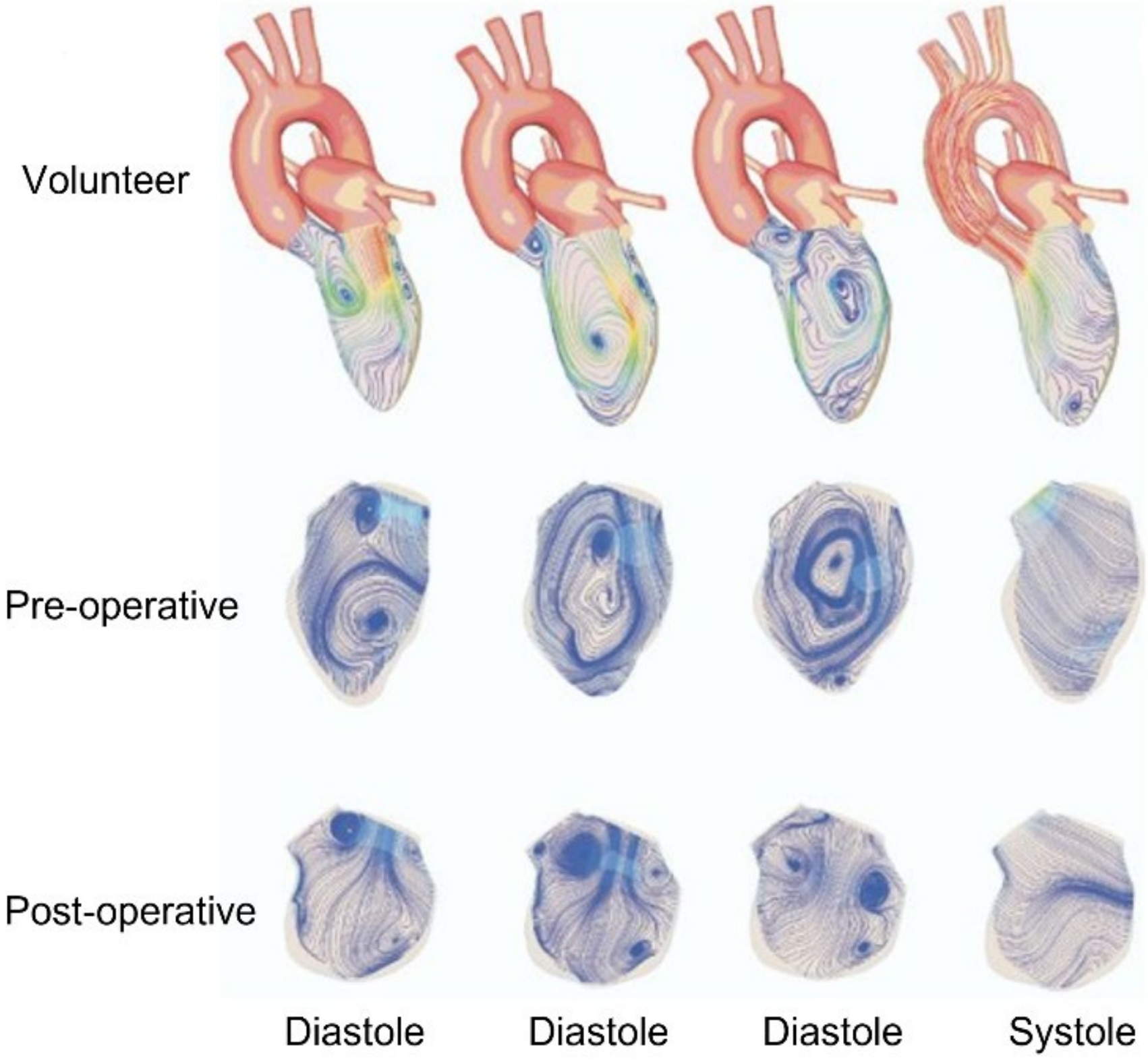

Fig. 6. Karlsruhe Heart Model (KHM) simulations showing comparison of vortex patterns in DCM models (pre-operative and post-operative) as opposed to a healthy model. Adapted with permission from previously published numerical simulations (Doenst *et al.* 2009).

As we have discussed so far there has been a systematic effort to study vortex dynamics in the left ventricle and its linkage to disease progression; also, a number of vortex indexes have been proposed. Table 1 shows the models, methods used, and limitations for various LV numerical studies, Table 2 shows the experimental (in vivo & in vitro) methods and imaging techniques used in past studies with major outcomes, and Table 3 presents the relevant numerical studies and their outcomes on whole heart and left ventricle hemodynamics. However, this approach is has not yet gained acceptance in clinical practice, since CFD modeling presents several limitations for various LV numerical studies, and more consistency in CFD outputs in various studies is desired to gain confidence for clinical practice. The consistency in results can be achieved by sharing models and methodology used in open-source platforms so that other researchers can utilize and replicate the results obtained.

Table 1. Vortex dynamics studies using numerical methods in either the whole heart or LV models.

| First Author and year | LV Model | Flow type | Viscosity model | Boundary condition | CFD Solver | Major limitations |
|---|---|---|---|---|---|---|
| (Fedele *et al.* 2023) | 3D PS heart | Turbulent | Newtonian | LP BC | In house code | Newtonian blood |
| (Korte *et al.* 2023) | 3D PSLV healthy | Turbulent | Newtonian | Pulsatile IP | Ansys Fluent | Absence of MV, AV, and LA chamber |
| (Bucelli *et al.* 2023) | 3D PSLV healthy | Turbulent | Newtonian | LP BC | In house code | Newtonian blood |
| (Bennati *et al.* 2023) | 3D PSLV MVR | Turbulent | Newtonian | Pulsatile IP | OS solver | Newtonian blood, absence of EPL |
| (He *et al.* 2022) | 3D PSLV healthy | Laminar | Newtonian | Pulsatile IP | Ansys Fluent | Absence of MV, AV, and LA chamber |
| (Grünwald *et al.* 2022) | 3D PSLV healthy | Turbulent | Non-Newtonian | Pulsatile IP | Ansys Fluent | Absence of MV, AV, and LA chamber |
| (Colorado-Cervantes *et al.* 2022) | 3D PSLV healthy | Laminar | Newtonian | Pulsatile IP | COMSOL | Absence of MV, AV, and LA chamber |
| (Collia *et al.* 2022) | 3D PSLV & ILV healthy | Laminar | Newtonian | Pulsatile IP | In house code | Absence of LA chamber |
| (Meschini *et al.* 2021) | 3D ILV, healthy | Laminar | Newtonian | Pulsatile IP | OS solver | Ideal model, absence of LA chamber |
| (Collia *et al.* 2021) | 3D PSLV healthy | Laminar | Newtonian | Pulsatile IP | In house code | Absence of LA chamber |
| (Viola *et al.* 2020) | 3D ILV, healthy | Laminar | Newtonian | Pulsatile IP | OS solver | Ideal model |
| (Lantz *et al.* 2019) | 3D PSLV, healthy | Laminar | Newtonian | Pulsatile IP | Ansys CFX | Absence of moving MV & AV |
| (Chan *et al.* 2019) | 3D ILV, healthy | Laminar | Newtonian | Pulsatile IP | COMSOL | Absence of moving MV & AV |
| (Meschini *et al.* 2018) | 3D ILV, healthy | Laminar | Non-Newtonian | Pulsatile IP | OS solver | Ideal model, absence of LA chamber |
| (Tagliabue *et al.* 2017a) | 3D ILV, healthy | Laminar | Newtonian | Pulsatile IP | OS solver | Ideal model & absence of LA chamber |
| (Gao *et al.* 2017) | 3D PSLV healthy | Laminar | Newtonian | Pulsatile IP | OS solver | Absence of moving AV |
| (Domenichini and Pedrizzetti 2016) | 3D ILV, healthy | Laminar | Newtonian | Pulsatile IP | Inhouse code | Ideal model, absence of moving MV & AV |
| (Liao *et al.* 2016) | 3D PSLV, DCM | Laminar | Newtonian | Constant Inlet | Ansys Fluent | Static LV & constant inlet. |
| (Vedula *et al.* 2015) | 3D PS left heart model | Laminar | Newtonian | Pulsatile IP | Inhouse code | Absence of moving MV |
| (Song and Borazjani 2015) | 3D PSLV, healthy | Laminar | Newtonian | Pulsatile IP | Inhouse code | Absence of MV, AV, & LA chamber |
| (Khalafvand *et al.* 2015) | 2D PSLV & MMV | Laminar | Newtonian | Pulsatile IP | Ansys Fluent | 2D geometry, absence of LA chamber & AV |
| (Seo *et al.* 2014) | 3D PSLV, PS | Laminar | Newtonian | Pulsatile IP | Inhouse code | Absence of AV & LA chamber |
| (Vedula *et al.* 2014) | 3D ILV & healthy | Laminar | Newtonian | Pulsatile IP | Inhouse code | Ideal model, Absence of MV and AV |
| (Chan *et al.* 2013a) | 2D ILV & DCM | Laminar | Newtonian | Pulsatile IP | COMSOL | Ideal 2D axisymmetric Newtonian |
| (Chan *et al.* 2013c) | 2D ILV & DCM | Laminar | Newtonian | Pulsatile IP | COMSOL | Ideal 2D axisymmetric Newtonian |
| (Mangual *et al.* 2013) | 3D PSLV healthy | Laminar | Newtonian | Pulsatile IP | Inhouse code | Absence of MV, AV, & LA chamber |
| (Zheng *et al.* 2012) | 3D ILV & healthy | Laminar | Newtonian | Pulsatile IP | Inhouse code | Ideal model, Absence of MV and AV |
| (Doenst *et al.* 2009) | 3D PS left heart model | Laminar | Cross-model | Pulsatile IP | Inhouse code | Absence of MV, AV, & PS BCs |
| (Schenkel *et al.* 2009) | 3D PSLV healthy | Laminar | Cross-model | Pulsatile IP | Star-CD | Absence of MV, AV, & LV chamber |
| (Loerakker *et al.* 2008) | 2D ILV & DCM | Laminar | Newtonian | Pulsatile IP | Inhouse code | Ideal 2D axisymmetric Newtonian |
| (Domenichini *et al.* 2007) | 3D ILV & healthy | Laminar | Newtonian | Pulsatile IP | Inhouse code | Ideal model, Absence of MV and AV |
| (Pedrizzetti and Domenichini 2005) | 3D ILV & healthy | Laminar | Newtonian | Pulsatile IP | Inhouse code | Ideal model, Absence of MV and AV |
| (Domenichini *et al.* 2005) | 3D ILV & healthy | Laminar | Newtonian | Pulsatile IP | Inhouse code | Ideal model, Absence of MV and AV |
| (Cheng *et al.* 2005) | 3D ILV & healthy | Laminar | Newtonian | Pulsatile IP | ADINA-FSI | Ideal model, Absence of MV and AV |
| (Baccani *et al.* 2002b) | 2D ILV & DCM | Laminar | Newtonian | Pulsatile IP | Inhouse code | 2D axisymmetric, Laminar, Newtonian |
| (Baccani *et al.* 2002a) | 2D ILV & healthy | Laminar | Newtonian | Pulsatile IP | Inhouse code | 2D axisymmetric, Laminar, Newtonian |
| (Vierendeels *et al.* 2000) | 2D ILV & healthy | Laminar | Newtonian | Pulsatile IP | Inhouse code | 2D axisymmetric, Laminar, Newtonian |

IP – inlet profile, DCM – dilated cardiomyopathy, ILV – ideal left ventricle, PSLV – patient-specific left ventricle, MV - mitral valve, AV - Aortic valve, PS – Patient-specific, BC – boundary condition, MMV – moving mitral valve, OS – open source, MVR – mitral valve regurgitation, EPL – electrophysiology, Ansys Fluent,(ANSYS, 2022) COMSOL.(COMSOL, 2018)

Table 2. Vortex dynamics studies using experimental techniques in the left ventricle models.

| First author & year | Experimental Technique | Imaging modality | Major outcomes |
|---|---|---|---|
| (Becker *et al.* 2023) | VFM (In vivo) | ECG | Growing heart undergoes a transition to an adult vortex pattern over first 2 years with higher loss |
| (Njoku *et al.* 2022) | 4D flow MRI (In vivo) | 4D MRI | 4D flow MRI is better for mitral flow analysis as compared with Doppler ECG |
| (Di Labbio *et al.* 2022) | D-PIV (In vitro) | High-speed camera | Proposed flow topologies using braid for healthy & diseased LV & can be a measure of outcomes |
| (Hu *et al.* 2022) | 4D flow CMR | 4D CMR | The space needed for optimal vortex formation is observed in children and adults between 6-18 years |
| (Gülan *et al.* 2022) | 4D flow MRI | 4D MRI | Low shear stresses in ventricle denotes higher adaptability to blood flow & increased compliance |
| (La Gerche *et al.* 2022) | S-T ECG (In vivo) | 3D ECG | Young athlete heart can generate greater outputs with lesser resting volume. |
| (Monosilio *et al.* 2022) | S-T ECG (In vivo) | 2D ECG | Patients with heart failure by reduced EF shows lower & misaligned hemodynamics forces |
| (Zhang *et al.* 2021) | ECG (In vivo & In Vitro) | ECG | LV disfunction had lower vorticity & strain in comparison with heathy group |
| (Yang *et al.* 2021) | VFM (In vivo) | CD ECG | DCM LV affects the direction of vortex rotation & can be a good indicator for detection |
| (Wiener *et al.* 2021) | PIV (In vitro) | ECG | Stenosed MV increases Transmitral gradients, but increased VED during exercise |
| (Adabifirouzjaei *et al.* 2021) | VFM (In vivo) | CD ECG | Degree of diastolic energy loss is directly proportional to level of Transmitral flow velocity |
| (Matsuura *et al.* 2019) | VFM (In vivo) | Doppler ECG | Peak E vorticity was strongly related with intra-ventricular pressure difference in healthy dogs LV |
| (Han *et al.* 2019) | VFM (In vivo) | CD ECG | Healthy LV shows apically directed flow & severe cases shows bidirectional flow with small vortices |
| (Berlot *et al.* 2019) | VFM (In vivo) | CD ECG | Apical intraventricular velocity gradient was lowest in control groups in comparison with DCM |
| (Labbio and Kadem 2018) | TR-PIV (In vitro) | High-speed camera | Diastolic vortex reversal was observed in LV & an increase in energy dissipation in AR severities |
| (Chan *et al.* 2017) | Adaptive vectors | PC-MRI | Flow propagation velocity is affected by jet direction & measurement location. |
| (Stugaard *et al.* 2015) | VFM (In vivo) | CD ECG | Diastolic energy loss increases in AR proportional to its severity & can be a marker for severity of AR |
| (Bermejo *et al.* 2014) | 2D Doppler color-In vivo | PC-MRI | Patients with DCM shows larger & stronger vortices as compared with healthy patients |
| (Fortini *et al.* 2013) | D-PIV (In vitro) | High-speed camera | The asymmetric single vortex is generated & its coherence is not broken by second filling. |
| (Eriksson *et al.* 2013) | 4D flow MRI (In vivo) | MRI & ECG | DCM results into altered diastolic flow routes and impaired preservation of inflow KE at pre-systole |
| (Abe *et al.* 2013) | E-PIV (In vivo) | ECG | The change in vortex strength is related with LV performance. |
| (Kheradvar *et al.* 2012) | ECG (In vivo) | ECG | Mitral annulus velocity was lower in abnormal cases. VFT varying in normal & abnormal cases |
| (Kheradvar *et al.* n.d.) | E-PIV & D-PIV (In vitro) | Camera & ECG | Flow patterns obtained using E-PIV & D-PIV are comparable & E-PIV limits small scale features |
| (Kheradvar and Gharib 2009) | PFS (In vitro) | Catheter & sensors | Transmitral thrust is maximum when VFT is in range of 4-5.5 & it can be an index of LV health |
| (Kheradvar and Gharib 2007) | D-PIV (In vitro) | High-speed camera | Magnitude of recoil is maximum during vortex pinch-off & VFT range of 3.5-4.5 |
| (Pierrakos and Vlachos 2006) | D-PIV (In vitro) | High-speed camera | The critical value of VFN for healthy LV is 6. Vortex formation depend on valve design. |

D-PIV – particle image velocimetry, E-PIV – Electrographic particle image velocimetry, TR – time-resolved, CD – color Doppler, S-T – speckle-tracking, ECG – echocardiography, VFM – vector flow mapping, PFS – pulse flow simulator, VFN – vortex formation number, VFT – vortex formation time, AR – aortic regurgitation, VED – viscous energy dissipation, EF – ejection fraction, CMR – cardiovascular magnetic resonance.

Table 3. Major contribution from recent CFD studies to vortex dynamics in patient-based whole heart or LV models and important outcomes.

| First author & year | Important outcomes |
|---|---|
| (Fedele *et al.* 2023) | Modeled electromechanical model of the whole human heart that considers both atrial and ventricular contraction with myocardial architecture. Model able to reproduce healthy cardiac functions. |
| (Bucelli *et al.* 2023) | Provides 3D representation of cardiac muscles and hemodynamics using FSI. They included cardiac electrophysiology, active and passive mechanics, and hemodynamics. |
| (Bennati *et al.* 2023) | Modeled left heart in presence of MVR with mitral valve movement and observed rise in regurgitant jets through the mitral orifice impinging against the atrial walls leads to increased WSS compared with healthy case. |
| (Korte *et al.* 2023) | Patient-specific simulations show lower systolic KE during exercise in comparison with rest. Precise segmentation of moving LV is crucial for better accuracy. |
| (Grünwald *et al.* 2022) | Modeled patient-specific wall movement in LV for SRV and healthy cases. They found out reduced value of VFT and total KE in SRV patients. Vortex circulation is limited to center of LV which leads to reduction in washout of SRV patients. |
| (Lantz *et al.* 2019) | Performed intracardiac blood flow simulations in LV by varying inlet velocity profiles at pulmonary veins. Large differences in flow patterns observed in LA and negligible differences for LV at various inflow profiles. |
| (Mao *et al.* 2017) | A fully coupled FSI approach is used for modeling MV and AV movement and its effect on hemodynamics. Comparison of FSI results with a LV model without valves shows significant difference in the flow field. FSI approach is close to physiological values. |
| (Gao *et al.* 2017) | Numerical simulations are performed for MV-LV model using non-linear soft tissue mechanics of MV. This methodology shows good agreement with physiological values such as aortic flow rate, end systole and end diastole ejection fractions. |
| (Vedula *et al.* 2015) | Investigated flow patterns in LA and its effect on LV hemodynamics using medical images. A comparison of the ventricle flow velocities between physiologic and simplified LA model shows difference of about 10% of peak mitral velocities. |
| (Mangual *et al.* 2013) | Comparative analysis of hemodynamics is performed between health and dilated LV models. Reduction of energetic efficiency and increase of flow stagnation is observed in dilated cardiomyopathy cases. |

FSI – fluid-structure interaction, MVR – mitral valve regurgitation, WSS – wall shear stress, KE – kinetic energy, LV – left ventricle, SRV – single right ventricle, VFT- vortex formation time, AV – aortic valve, MV – mitral valve.

## 3.0 Challenges and opportunities in CFD modeling of the patient-specific left ventricle

The typical pipeline for patient-based modeling blood flow in the left ventricle is sketched in Fig. 7. This pipeline proceeds through stages, each of them with its own complications. The first stage is to collect the medical images (CT or MRI) of heart models from physicians. The images need to be segmented for extraction of heart surface and further smoothing. The extracted model is fed to either open-source or commercial tools for mesh generation and simulation. Mesh independence tests are crucial before performing final simulations. Post mesh independence assessment, the appropriate patient-specific boundary conditions are enforced for heart model such as inlet flow/pressure profiles and ventricle wall movement. The blood flow dynamics in the heart model is simulated by solving numerically the Navier-Stokes equations. After solving the flow field, the three-dimensional vortical structures and various hemodynamic parameters can be extracted for flow analysis. The most used marker for heart health are the vortex structures and vortex formation number. These markers can be helpful to monitor the heart health and can also be beneficial for heart surgery planning or evaluation of post-surgery outcomes. Every step involved in pipeline of blood flow modeling is further discussed in depth as follow:

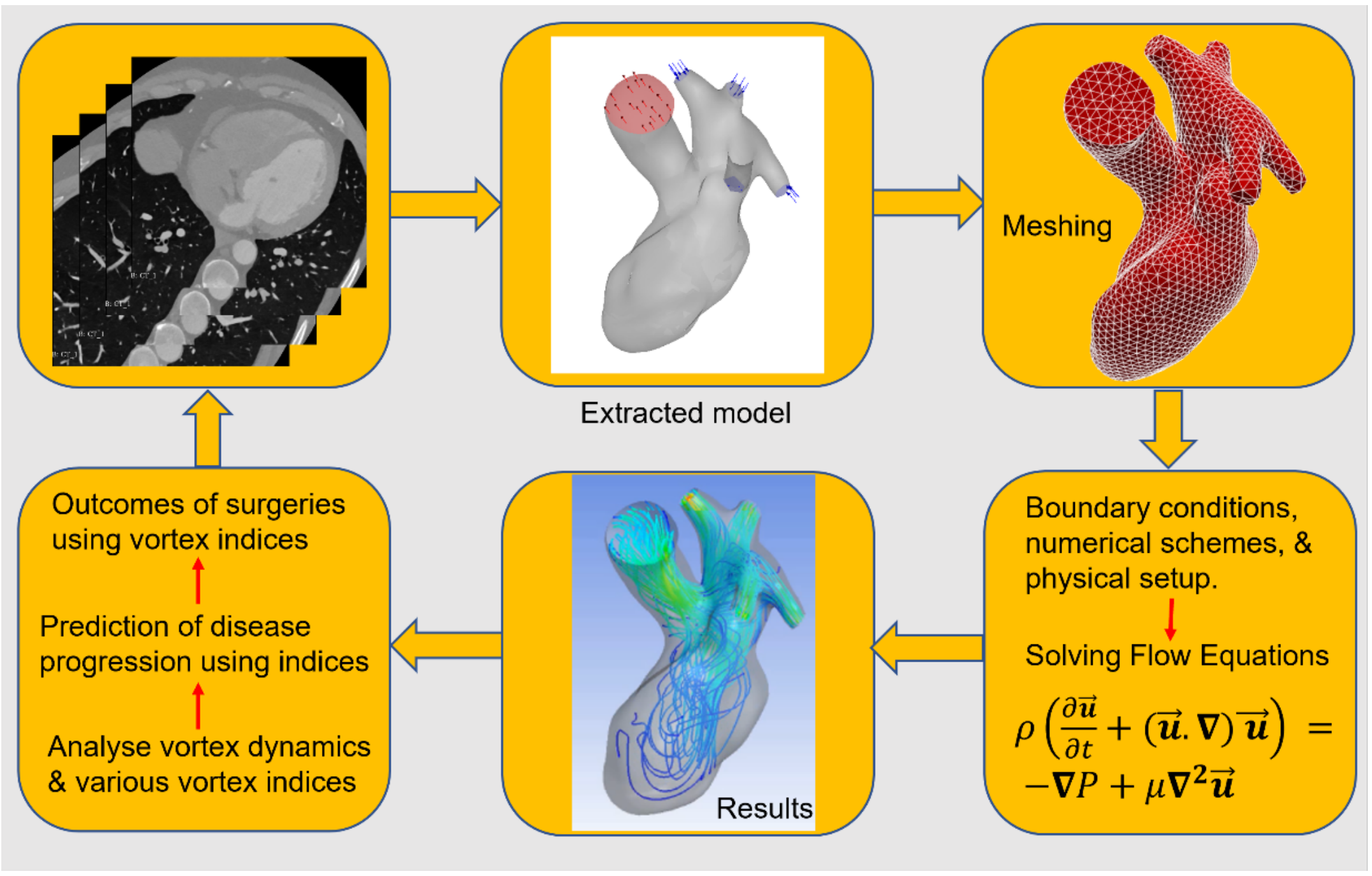

Fig. 7. Pipeline used in CFD modeling of vortex dynamics in patient-specific left heart model.

### *3.1 Segmentation of heart models using medical images*

The DICOM file format is commonly used in medical images for visualization, transfer, and segmentation purposes. The segmented models can be exported in stereolithography (STL) file formats for further processing, refinement, and numerical simulations. The segmentation step involves generation of the endocardial surface and definition of endocardial displacement field during systolic filling dilation. Plenty of tools are available for segmentation of medical images. However, most popular tools are discussed in this section. The entire segmentation procedure can be performed using open-source software 3D slicer (Kikinis et al. 2013), the vascular modeling toolkit (Antiga *et al.* 2008), vmtk features can be enhanced by using additional tools for cardiac surface processing and the SimpleElastix library for image processing (Fedele and Quarteroni 2021; Klein *et al.* 2010). The segmented and processed models can be visualized using ParaView, an open-source visualization software (A. Henderson 2007). Although, above mentioned libraries are available, the segmentation and generation of whole heart model geometry with its moving endocardial walls and heart valves is cumbersome and challenging task.

The accurate segmentation and reconstruction of heart models and its time-resolved wall motion using cine-MR or ultrasound images enhances the diagnostic accuracy and CFD modeling credibility. However, due to limitation of imaging resolution, lack of accuracy in imaging of heart valves and ventricular apex is observed. The wall motion in very short time interval during cardiac cycle is difficult to resolve properly due to limitation of temporal resolution in imaging modalities. The number of frames per cardiac cycle is limited to 20-30. The intermediate surfaces between very short intervals of time are obtained by interpolation of temporal datasets (Antiga *et al.* 2008; Renzi *et al.* 2023).

Regarding the segmentation techniques, two main approaches exist: image-driven approaches and model-driven approaches. Image-driven models are designed as without or with weak prior models, while model-driven approaches are based on strong prior knowledge. The image driven approach identifies the pixel difference in tissue, blood pool, and myocardium and segmentation can be performed by thresholding and region growing methods (Peng *et al.* 2016). On the other side, model-driven approach uses strong prior knowledge about specific shape variability of the ventricle chambers, instead of making assumptions on boundaries. Various researchers have used model-based approach for segmentation of heart and this method has shown promising results (Mitchell *et al.* 2002; van Assen *et al.* 2006).

Choice of segmentation techniques is not trivial and can be constrained by the specific protocol. Model-based technique can be useful for obtaining LV walls combined with thresholding to eliminate the effect of papillary muscles. If training datasets are limited, then the use of model-based techniques is not preferred, and it is instead advised to refer to image-driven approaches. Finally, segmentation accuracy is the major criteria for selecting segmentation technique.

Challenges of whole heart segmentation are mainly due to large shape change of the heart and unclear boundaries between various substructures. A practical problem is the computational time for segmentation raised from nonrigid registration process. Other challenges that arise in use of fully automatic algorithms are large variability of the heart shape, indistinct boundaries, and restricted image quality or resolution. The automatic algorithms cannot be applied to every heart simulation, but rather they are limited to specific tasks.

### *3.2 Meshing and grid independency assessment*

Pre-processing and mesh generation is a highly human-labor-intensive modeling stage, typically involving repetitive tasks. Many times, the quality of the refined mesh is uncertain. The meshing of the heart models can be performed by various open-source and commercial codes. Most of the studies used unstructured meshes which are computationally expensive in terms of convergence time per iteration. The structured mesh topology, whenever applicable, will be a good choice for fast convergence and low computational time. For the same accuracy of result, six times less cells and fourteen times less computational time are required for structured mesh in cardiovascular flow simulations (De Santis *et al.* 2010). Note however that structured mesh generation in patient-based left ventricle is a tedious and time-consuming task. However, in past, structured mesh has been used for hemodynamics in curved and geometrically complex aortic artery and its sub junctions (Manchester *et al.* 2021). Recently, researchers have developed machine learning algorithms for creating automatic segmentation and meshing of arterial and heart models (De Santis *et al.* 2010; Kong *et al.* 2021). Some of the commonly used automated algorithms are pyFormex (pyFormex 2023), a python-based open-source software dedicated to create structured mesh for patient-specific left coronary tree. Similarly, the deep learning model MeshDeformNet has been developed for creating deforming meshes with great accuracy using CT and MR image data for whole heart model (Kong *et al.* 2021). The MeshDeformNet model can efficiently construct 4D whole heart dynamics that captures the motion of a beating heart from a time-series of images. The number of surface meshes and their connectivity must match at various time frames in moving mesh methodology setup.

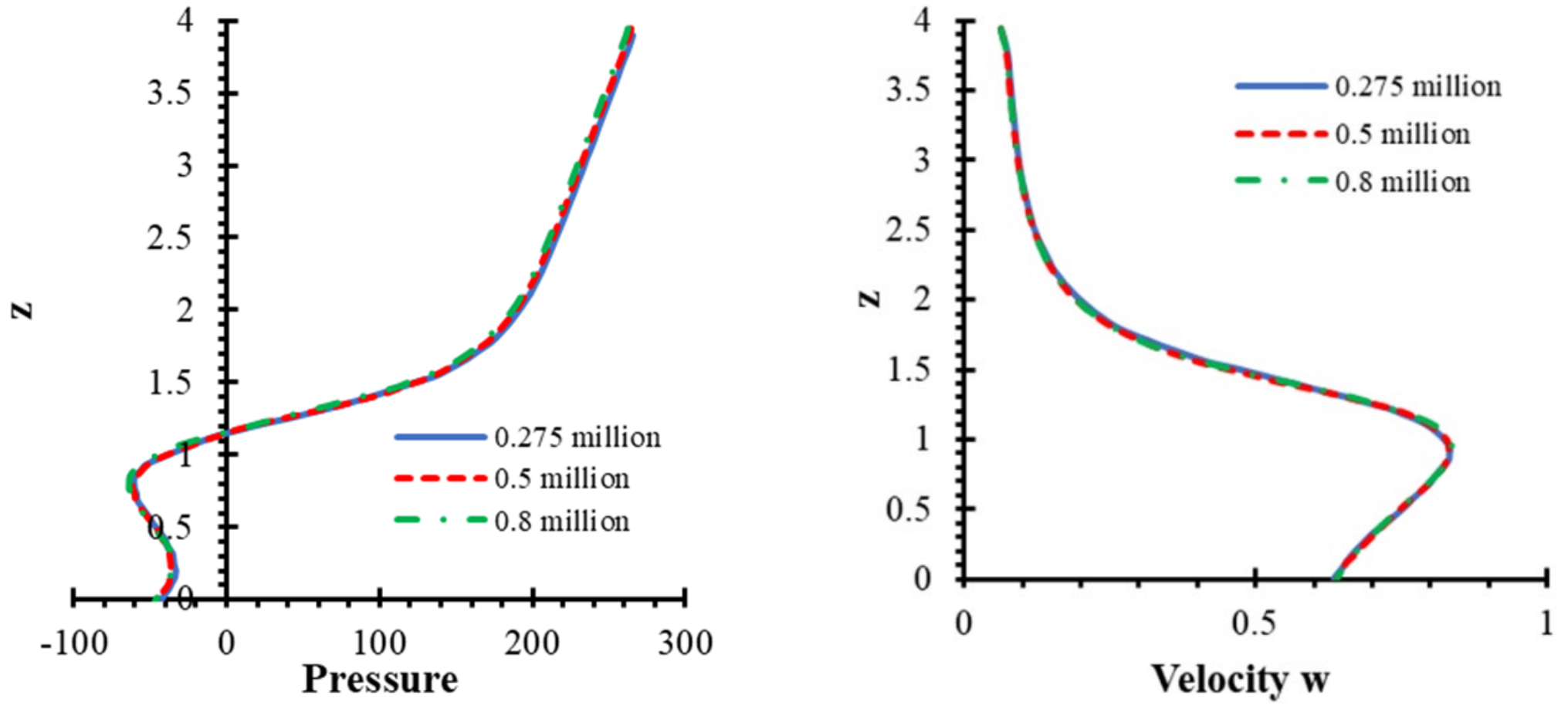

Figure. 8. Mesh independence study in an ideal left ventricle model (a) pressure profile (b) axial velocity profile at central axis of mitral orifice (inlet) plotted at various mesh densities.

Mesh independence tests are essential to obtain a credible result. Several important flow parameters are compared at various mesh densities and after certain mesh refinement, no change or bounded change in flow parameters should be observed. The refined meshes are used for further simulations and analysis of results. The refined meshes are required for the accurate prediction of the complex flow patterns downstream of the heart valves (Ge *et al.* 2003). In cardiovascular fluid mechanics, the commonly used flow parameters for assessing mesh independence are wall shear, velocity profiles, kinetic energy, and pressure profiles (Nagargoje and Gupta 2020a, 2020b; Nagargoje *et al.* 2021, 2022, 2023). Figure 8 shows the mesh independence study for an idealized left ventricle (Fig. 9a) numerical simulation. The difference between pressure and velocity values between 0.5 million and 0.8 million elements are below 0.5%. Therefore, the 0.5 million elements are used for further simulations and numerical analysis.

### *3.3 Numerical Schemes Consistency*

We observed a disparity in various numerical schemes used for modeling of left ventricle hemodynamics. Few studies used first order temporal or spatial discretization, while others have used second or higher order discretization methods. Similarly, different pressure-velocity coupling schemes has been used in past studies. It is not clear about which order of discretization or pressure-velocity coupling is best for biomedical simulations. Sufficient knowledge on various numerical schemes and its effect on vortex patterns in the left ventricle would be a great guide for *in silico* medicine. However, most of the studies used second order schemes for temporal discretization and second or higher order schemes for spatial discretization. Nowadays, use of higher order schemes in industrial applications is a common practice. The appropriate choices on use of numerical schemes with the in vivo or in vitro validation can be an excellent addition to available literature knowledge. The universal guide for CFD methods used and consistency in results obtained could be a better approach to translate CFD results into clinical decisions. The successful adaption of CFD to clinical decision making can be possible when CFD output is readily tuned for specific clinical needs.

### *3.4 Boundary conditions specification*

The effect of boundary conditions on flow patterns and hemodynamic markers have been thoroughly studied and shows the necessity of using individual patient-specific boundary conditions (Dahl *et al.* 2012; Lantz *et al.* 2019). Numerical results significantly depend on the accuracy of boundary conditions. Inappropriate boundary conditions lead to incorrect prediction of flow fields and hemodynamic markers (Tagliabue *et al.* 2017b). Use of generic boundary conditions available in commercial solvers restricts the use of CFD analysis in medical practice due to deviations from in vivo physiology. However, a recent study explored the effect of various inlet boundary conditions on hemodynamics and advised that artificial inlet profiles are acceptable to use in absence of patient-specific boundary conditions (Wei *et al.* 2019). Artificial inlet profiles are generic inlet boundary condition adapted from literature studies which consists of E-wave and A-wave (diastolic filling), as shown in Fig. 4 and 9(b). The inlet profile is similar in all healthy cases, but it can be different in diseased cases (Wei *et al.* 2019). Patient-specific inlet boundary conditions are replication of in vivo velocity/flow profiles measured by phase-contrast MRI or catheterization. Using such inlet profiles in CFD simulation for each patient-specific left ventricle replicates the CFD results close to in vivo flow patterns. The flow profiles obtained from medical data can be implemented in CFD simulations by interpolating the flow curve and writing a customized function (e.g., in Ansys, a User-Defined Function) for boundary condition implementation. Note that obtaining the blood flow velocities or pressure at various valve locations is not a routine clinical practice, which limits the availability of patient-specific boundary conditions and accurate flow dynamics in the heart. It can be a very good initiative to perform whole heart simulations using patient-specific flow and wall properties. Such studies will be a valuable piece of knowledge to existing literature, and it will be possible to translate such studies in clinical practices.

We have simulated and compared the effect of plug and parabolic inlet profile on vortex formation and found that vortex patterns are significantly different, as shown in Fig. 9(d). The vortex patterns obtained using plug flow inlet is in agreement with in vivo flow dynamics and past published articles (Chan *et al.* 2013d). Therefore, it is recommended to use plug inlet profile at left ventricle inlet (mitral valve) during left ventricle filling simulations. In a recent study, three different models of left heart models with varying inlet pulmonary veins (PVs) were compared and analyzed (Dahl *et al.* 2012). Four jets enter the atrium asymmetrically, and complex vortex patterns are observed. They found that with anatomically based PVs positions, the flow directed towards MV without collision. This model shows evenly distributed velocity at MV plane and lower maximum transmitral velocity during E-wave. The asymmetrically located veins prevent flow instabilities and excessive energy dissipation in the flow. To obtain physiologically correct simulations, ventricular filling and mitral valve dynamics should be modeled using patient-specific anatomies, flow rates, and mitral valve properties. In other recent study, three different inlet waveforms at pulmonary veins are analyzed using 4D flow MRI (Lantz *et al.* 2019). They have observed that different inlet at PVs affect the left atrium flow patterns. Large variation in kinetic energy (KE) is observed for varying inlet velocity profiles, especially during early filling phase. In vivo MRI measurement shows higher flow volumes on the right side of PVs compared with left side, which seems to be realistic as the right lung has three lobes and left lungs has only two lobes. Some of the vortical structures generated in the left atrium (LA) transferred towards LV. The high residence time regions trigger thrombus formation and it is most commonly observed at abnormal flow regions in LA (Hara *et al.* 2009; Heppell *et al.* 1997). Asymmetrical filling of the LA preserves the momentum and redirected towards LV through mitral valve (Kilner *et al.* 2000).

### *3.5 Wall modeling*

Moving mesh methods are useful for simulating the large expansion characterizing the movement of left ventricle walls. A possible implementation of moving mesh approach interprets interpolating left ventricle surface generated at various time instances during a cardiac cycle. The change in left ventricle volume is reconstructed from medical images in spatiotemporal node positions. The mesh is updated to respective dilated left ventricle volume using surface coordinates at various time instances. It is very difficult to use two-way fluid-structure interaction (FSI) for larger deformations of the left ventricle. However, FSI is useful for modeling heart valve movement. In the FSI, the momentum transfer from fluid (blood) to solid (artery wall/endothelial surface) and vice versa is modeled. Current spatial and temporal resolution for cardiac MRI is around 1-1.5 mm /40-50 ms, respectively (Saeed *et al.* 2015). This resolution is inadequate to capture the motion of heart valves. Very few studies have modeled patient-specific heart models with heart valve movement (Chnafa *et al.* 2014; Gao *et al.* 2017; Mao *et al.* 2017; Seo *et al.* 2014; Su *et al.* 2014, 2016). In most of the studies, valves are modeled as either fully open or fully closed; moreover, valve opening and closing is assessed to be very fast, almost instantaneously. However, neglecting wall movement during opening and closing may affect the left ventricular flow patterns. Valve movement has been modeled in two ways, either using medical imaging data to specify opening and closing motion or directly simulating using FSI approach. It is necessary to give wall boundary conditions during early diastole to aortic valve and during systole to mitral valve. Specifying wall boundary condition defines the valve as completely closed. These can be changed during a cardiac cycle from wall to velocity inlet or pressure outlet and vice versa. Specifying these boundary conditions replicates the opening and closing of valves during diastole and systole phase of cardiac cycle. Despite many investigations having been conducted by using different types of boundary conditions to LV wall, it is still unclear about the best fit of boundary conditions to replicate physiological flow features in LV.

### *3.6 Newtonian and non-Newtonian blood models*

It has been argued that blood flow in larger arteries and veins can be modeled as Newtonian (Ku 1997). The heart is, in a sense, a large vessel, but it involves complex flow patterns due to contraction and expansion of endocardial walls and moving heart valves. The flow patterns in the heart chamber involve vortices and recirculation. The recirculation leads to lower shear rate values, and it is well-known that blood viscosity is dependent on local shear rates (Cho and Kensey 1991). The lower shear rate leads to increased blood viscosity, and it becomes important to understand the effect of non-Newtonian models on vortex patterns in LV. A recent study attempted to investigate the effect of various non-Newtonian models (shear-thinning) on vortex analysis in the LV and they found a significant difference between both vortex patterns and vortex ring angle in various non-Newtonian models (Doost *et al.* 2016b). The number of smaller vortices and their magnitudes are different for various non-Newtonian models. A larger apparent viscosity value is observed in apex and middle of left ventricle. The vortex patterns are the major marker of measuring heart function, and it shows the importance of non-Newtonian blood flow modeling in LV. They have analyzed the effect of non-Newtonian blood for a single case. It would be interesting to see comparison of non-Newtonian models in large datasets of patient-specific left ventricles, especially in dilated cardiomyopathy (DCM) patients. In DCM patients, left ventricle volume is increased and larger stagnation regions are observed this can lead to increased viscosity regions. To the best of authors knowledge, there is no study available on effect of vortex formation number (VFN) for various non-Newtonian models in healthy and diseased LVs. The VFN number has been considered as an important marker of identifying normal and abnormal vortex patterns in LV. Overall, the consideration of non-Newtonian modeling becomes important due to its strong linkage with vortex patterns and vortex formation number.

Few other studies have investigated the effect of non-Newtonian models on hemodynamics of mechanical heart valves and Left Ventricular Assist Device (LVAD) (Al-Azawy *et al.* 2017; De Vita *et al.* 2016) . They found that irrespective of having larger vessels such as aorta and heart pump, shear rate values are lower than threshold limits to show constant viscosity and its shear-thinning behavior might affect the flow patterns. Hemolysis induced due to altered wall shear stress (WSS) is observed to be higher for non-Newtonian model compared with Newtonian blood (De Vita *et al.* 2016). Further, it was suggested to use non-Newtonian models for modeling left ventricular assisted device. They found that the shear rate values are lower than 100 $s^{-1}$ in the chamber and these shear rate values belong to nonlinear viscosity range for blood (Al-Azawy *et al.* 2017). Total kinetic energy is observed to be higher for non-Newtonian models compared with Newtonian one. Finally, there is insufficient knowledge available on which non-Newtonian model is best fit for hemodynamics in left ventricle (De Vita *et al.* 2016).

### *3.7 Validation of numerical methodology*

Validation of numerical results against well-designed experimental datasets is necessary for gaining confidence in CFD simulations. The accuracy of CFD simulations in biomedical flows largely depends on selection of geometry, boundary conditions, blood and wall properties, mesh quality, and numerical methods used. Due to simplifications and assumptions made during heart simulations, it is very difficult to validate these results with experimental or in vivo data. Very few studies validated their CFD results against the experimental values. Vedula et al. have validated hemodynamics in a moving left ventricle model with particle image velocimetry experiments, which matched well qualitatively and quantitatively (Vedula *et al.* 2014). Few other studies validated CFD results with in vivo flow profiles obtained using 4D flow MRI and they matched qualitatively (Saber *et al.* 2003; Schenkel *et al.* 2009).

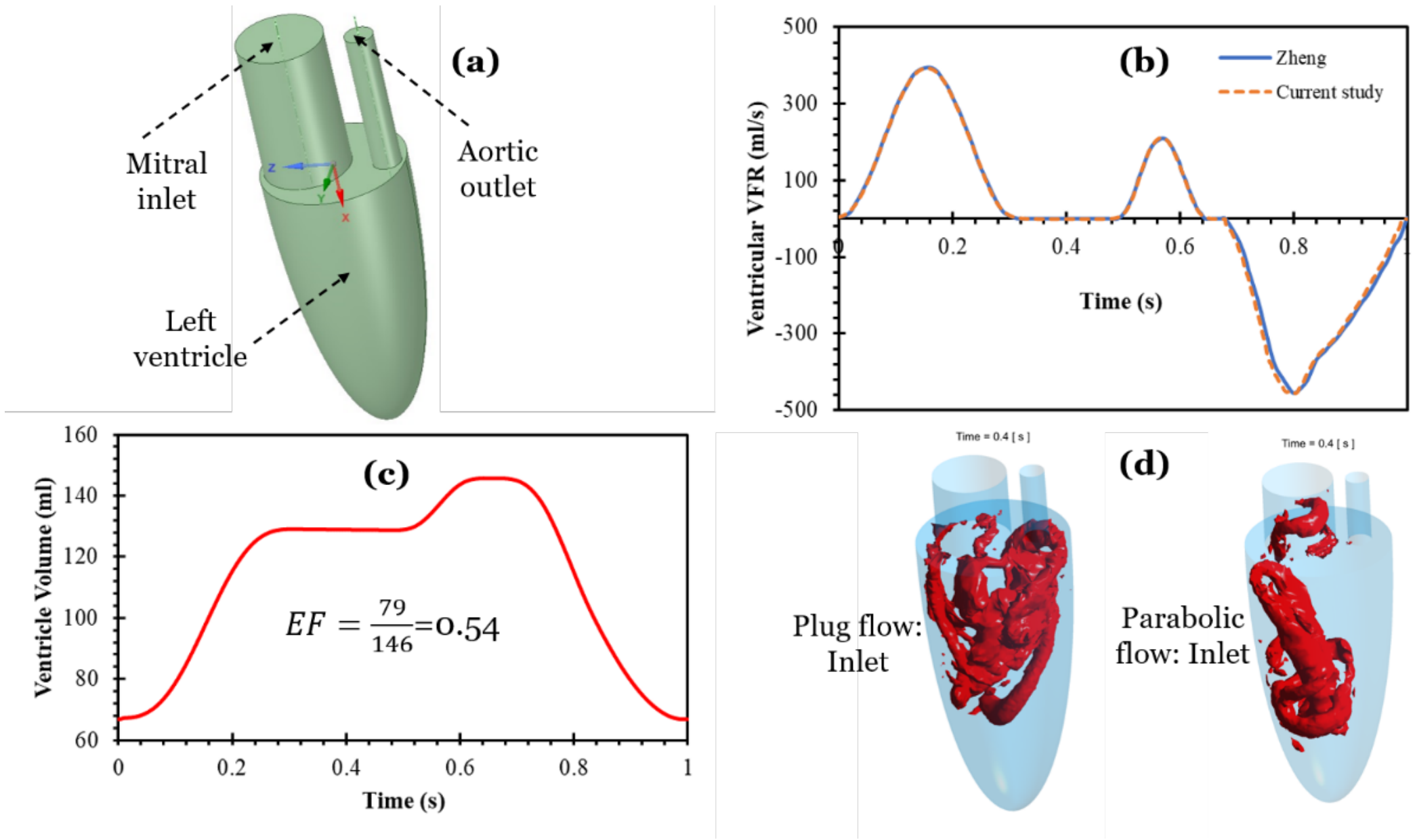

Fig. 9. Validation and comparison of inlet flow profiles with Zheng et al. (2012)(Zheng *et al.* 2012) using moving mesh methodology: (a) Ideal geometry of left ventricle, (b) Validation of ventricular volumetric flow rate (VFR) with past study, (c) Ventricle volume change during cardiac cycle, (d) Effect of plug and parabolic inlet profile on Q-criterion in ideal left ventricle model.

We have validated numerical methodology with Zheng et al. on ideal left ventricle hemodynamics, as shown in Fig. 9 (Zheng *et al.* 2012). Simulations were performed using well established moving mesh approach. The movement of heart walls during diastolic filling have been mimicked by updating surface files in a short interval of time during simulations, please see section 3.4 for more detailed explanation. Figure 9(a) shows the ideal left ventricle model (Vedula *et al.* 2014), which is a semi-prolate ellipsoid in shape consisting of mitral orifice (inlet) and aortic outlet. The inlet profile is adapted from literature and compared the Volumetric Flow Rate (VFR) at inlet and outlets and it has already been discussed in the boundary condition section. Figure 9(b, c) shows the validation of the ventricular VFR for present simulation with Zheng et al., and ejection fraction (ratio of ejected volume to the total LV volume) achieved is matching with their results (Zheng *et al.* 2012).

### *3.8 New tools for pattern identification*

The identification of coherent structures is of paramount importance in the field of cardiac flow modeling. However, this area remains under-explored due to the inherent complexity of the problem. New data-driven techniques offer a promising avenue for advancing our understanding of flow physics, providing general descriptions of the main mechanisms involved in heart dynamics, and to detect the presence and evolution of the vortex ring originated in early diastole from the mitral valve.

For instance, data-driven modal decomposition techniques can identify the primary patterns and global instabilities driving flow dynamics. Two prominent methods in fluid dynamics are proper orthogonal decomposition (POD) and dynamic mode decomposition (DMD). POD decomposes the flow into the most energetic orthogonal modes, representing large-scale structures. DMD identifies high-amplitude modes driving flow dynamics. By identifying the main flow patterns, it is possible to connect these structures with healthy or diseased heart mechanisms, allowing for the preemptive detection of cardiovascular diseases. These data analysis techniques have gained traction in medical image analysis and diagnosis(Fathi *et al.* 2018; Grinberg *et al.* 2009), extending to conditions like Parkinson's and lung disease(Fu *et al.* 2020; Xi and Zhao 2019).

In particular, a study by Groun et al. has investigated the use of higher-order dynamic mode decomposition (HODMD) in the realm of medical imaging (Groun *et al.* 2022), particularly in the analysis of echocardiography images obtained from mice with different cardiac conditions. In this work, the algorithm demonstrated robust performance, successfully capturing two branches of frequencies related to heart and respiratory rates across all datasets. Despite variations in the number of identified modes and frequencies due to factors such as noise levels, disease characteristics, and anesthesia effects, HODMD consistently distinguishes characteristic patterns associated with each cardiac pathology.

On this section, we want to explore the potential of HODMD in analyzing cardiac flow CFD databases, shedding light on vortical structures within the LV model. Assume that temporally resolved data fields like pressure and velocity are extracted for each time instant from our CFD simulation of the left ventricle. This data is compiled into a three-dimensional snapshot tensor called $\boldsymbol{v}(x, y, z, t_k)$, which we feed into the HODMD algorithm. This method decomposes the dataset into an expansion of DMD modes $\boldsymbol{u}_m$. Each of them is associated with an amplitude $a_m$, as

$$\boldsymbol{v}(x, y, z, t_k) \simeq \sum_{m=1}^{M} a_m \, \boldsymbol{u}_m(x, y, z) e^{(\delta_m + i\omega_m) t_k},$$

for $k = 1, \ldots, K$ . The DMD modes can grow, decay, or remain neutral according to the growth-rate $\delta_m$ associated and oscillate in time with frequency $\omega_m$.

Notably, this algorithm exhibits several advantages, including its capability to analyze three-dimensional data, integrate both experimental and numerical data sources, and obtain a set of modes associated with the vortical structures.

Recent advances in deep learning, including convolutional neural networks (CNNs) and recurrent neural networks (RNNs), have significantly enhanced the ability to manage and interpret the large datasets generated by improved cardiac flow simulations. These AI and machine learning techniques have been instrumental in diagnosing cardiovascular diseases, particularly through the classification of MRI image sequences of coronary arteries (Berikol *et al.* 2016; Worden *et al.* 2015).

In particular, the work by Bell-Navas et al. introduces an automatic cardiac pathology recognition system using a novel deep-learning framework (Bell-Navas *et al.* 2024). This system analyzes real-time echocardiography video sequences in two stages. The first stage converts echocardiography data into annotated images suitable for machine learning, employing the HODMD algorithm for data augmentation and feature extraction. The second stage involves training a Vision Transformer (ViT) from scratch, adapted for small datasets. This neural network predicts heart conditions from echocardiography images and has shown superior performance, even surpassing pretrained CNNs, highlighting the efficacy of the HODMD algorithm in the medical field.

In conclusion, the application of the HODMD algorithm to the analysis of cardiac flows has not only the potential to deepen our insight into the intricate physics governing this complex phenomenon but also may be of use as a valuable tool for identifying and foreseeing the temporal progression of cardiovascular diseases. Through its ability to unravel the flow patterns within the left ventricle, HODMD illuminates the path to more precise and targeted interventions.

## 4.0 Future directions

A recent study modeled the whole beating heart hemodynamics with inclusion of valves and electrophysiology (Fedele *et al.* 2023). However, we are still far away from replicating the physiological phenomena. It is recommended to use following advancements to move closer towards replication of in vivo flow and material properties:

***4.1 Imaging and segmentation:*** Current limitation in temporal and spatial resolution of imaging techniques restricts the acquisition of heart valve anatomies, myocardium thickness and structures, papillary muscles, and *chordae tendineae*. Inclusion of these anatomical features in vortex analysis would give a more accurate and closer picture of in vivo flow patterns. The inclusion of papillary muscles may change the vortex patterns by obstructing blood flow stream and vortices. Segmentation of such minute anatomic features is a major challenge. One possible avenue to overcome this obstruction would be developing neutral network algorithms for segmentation. Machine learning-based algorithms could be explored then to interpolate these structures from available weak anatomical features.

***4.2 Numerical simulation:*** The wall movement of left ventricle has been modeled so far by dynamic meshing features in each available study. However, the surface meshes used and their respective time instance during cardiac cycle must match to accurately predict the flow patterns in LV. Interpolation techniques are used for intermediate surface meshes during transient simulations. It is very difficult to interpolate the surface movement by including the moving heart valves and papillary muscles. Also, patient-specific boundary conditions are needed for inlet velocity/pressure at local regions, and it can be measured using probes in each patient. The aortic and mitral valve properties can be replicated using heart muscle stress test, and this could be implemented into numerical simulations. The computational time and cost of cardiovascular simulations is decreasing rapidly, and it can be extended by parallelizing the flow solvers. The cost of computer simulation per Gflop is less than a cent of dollar, which can be utilized in patient care by including numerical simulations in medical practice.

***4.3 Analysis of results using trained models:*** The cardiovascular simulations are time consuming due to human interventions in every step and the time required can be reduced by automating the numerical simulations from loading images to post-processing of results. The vortex patterns in LV can be analyzed by machine learning trained algorithms on various scales of healthy, moderate, to severe diseases. After gaining large and rich CFD data on vortex dynamics for large population, it could be possible to train neural networks and obtain vortex parameters using just an anatomic model of heart.

The inclusion of the above suggestions will significantly contribute to increase the applicability of CFD simulations in daily clinical practice and pre-operate and post-operate surgery planning and evaluation. Use of increased computational capability and lower cost will enhance the inclusion of in silico medicine in clinical practice and decision-making. The US Food and Drug Administration (FDA) recommends using computer simulations to complement human or animal testing. Recently, the FDA approved multiple computer programs that are useful in the treatment and management of cardiovascular diseases (Ahmed *et al.* 2023; Morrison *et al.* 2018).

## Declaration of Competing Interest

The authors declare that they have no known competing financial interests or personal relationships that could have appeared to influence the work reported in this paper.

## Funding statement

The authors acknowledge the grants PID2020-114173RB-I00, TED2021- 129774B-C21 and PLEC2022-009235 funded by MCIN/AEI/ 10.13039/501100011033 and by the European Union “NextGenerationEU”/PRTR, and S.L.C. acknowledges the support of Comunidad de Madrid through the call Research Grants for Young Investigators from Universidad Politecnica de Madrid.

**Data availability statement.** Not applicable.